\documentclass[
reprint,  floatfix,
amsmath,amssymb,aps,showpacs,superscriptaddress,dvipdfm   %
nofootinbib, showkeys
]{revtex4-1}
\usepackage{amssymb}
\usepackage{graphicx}
\usepackage{booktabs}
\usepackage{dcolumn}
\usepackage{bm}
\usepackage{mathrsfs}
\usepackage{mathtools}
\usepackage{amsmath}
\usepackage{amssymb}
\usepackage{epstopdf}
\usepackage{units}
\usepackage{mathtools}
\usepackage[mathlines]{lineno}
\usepackage{dsfont}
\usepackage{upgreek}
\usepackage{subfig}

\begin{document}


\title{The Geometry of Most Probable Trajectories in Noise-Driven Dynamical Systems}

\author{John C. Neu}
\affiliation{Duke University, Department of Biomedical Engineering, Box 
90281
Durham, NC 27708-0281}

\author{Akhil Ghanta}
\affiliation{Duke University, Department of Physics, Box 90305
Durham, NC 27708-0305}

\author{Stephen Teitsworth}
\affiliation{Duke University, Department of Physics, Box 90305
Durham, NC 27708-0305}

\date{\today}

\begin{abstract}
This paper presents a heuristic derivation of a geometric minimum action method that can be used to 
determine most-probable transition paths in noise-driven dynamical systems. Particular attention is 
focused on systems that violate detailed balance, and the role of the stochastic vorticity tensor is 
emphasized. The general method is explored through a detailed study of a two-dimensional quadratic 
shear flow which exhibits bifurcating most-probable transition pathways.

\end{abstract}

\pacs{}
\keywords{transition path, geometric stochastic action, fluctuation loops}
\maketitle


\section{Introduction}
\label{sec:1}
In the final chapter of Feynman and Hibbs' classic text on path integrals \cite{Feynman_1965}, the 
scattering of a fast particle from nucleii in a slab of material is examined in order to answer the 
question:  what is the most probable path of the particle from emitter to detector? The relative 
probabilities of noise sequences which produce paths with the required endpoints can be represented 
as a functional of the paths themselves.  The variational characterization of the most probable path is 
closely analogous to least action paths in mechanics, or geodesic paths in geometry (a special case of the 
former).  Independently, a rigorous theory of most probable paths in stochastic dynamical systems was 
developed by Wentzell and Freidlin \cite{WentzellFreidlin_1970, WentzellFreidlin_2012}, and 
subsequently elaborated and explored by many others, see e.g. \cite{Berglund_2006}.  Additionally, a 
similar formalism for treating large noise-induced deviations has been applied more recently to 
spatially-extended hydrodynamic models \cite{Prados_PRL_2011, Bertini_RMP_2015}.

In applications to concrete problems, physicists \cite{Maier_SIAM_1997, Luchinsky_RPP_1998} often 
employ the Legendre transformation to convert the Lagrangian formulation of most probable paths into 
an equivalent Hamiltonian formulation.  Here we retain the Lagrangian formulation, to see very directly 
some geometric aspects of most probable paths (see Sec. II).  A geometrical characterization of most 
probable paths was recently introduced by Heymann and Vanden-Eijnden \cite{Heymann_PRL_2008, 
Heymann_CPAM_2008, Heymann_2015}, and we recapitulate aspects of their treatment in the language 
of classical mechanics at the level of a typical physics graduate course. A chief advantage is that there 
are very clear expressions of some common insights:  in the small noise limit, almost all paths with fixed 
endpoints closely follow the most probable path in an almost deterministic manner.  Furthermore, the 
speed along the most probable path is the deterministic speed.  As the noise decreases, the transit time 
between endpoints remains fixed, but the expected time between successful transits becomes 
exponentially large.  

In Section III, we discuss the geometric meaning of \textit{detailed balance}, a fundamental notion from 
statistical physics. The traditional meaning of detailed balance, or its breaking, is framed in terms of the 
probability current in the Fokker-Planck equation:  ``fence off" a bounded region of state space with the 
impermeable boundary condition, and allow the probability density to relax to a steady state within that 
region.   For a detailed balance system, the probability current vanishes identically, and there is a 
potential energy function so that the equilibrium probability density is proportional to a Boltzmann 
factor with this potential.  If detailed balance is broken, the steady probability current is divergence-
free, as it must be, but does \textit{not} vanish identically \cite{Maier_PRL_1993, Zia_JStatMech_2007}.  
We review how this traditional characterization of detailed balance restricts the velocity field and noise 
tensor of the stochastic dynamics, so that the \textit{stochastic vorticity tensor} vanishes. Furthermore, 
there is a connection to the geometry of most probable paths:  in detailed balance systems, the 
``backward" path associated with an interchange of the starting and destination endpoints is the 
reversal of the original ``forward" path.  If detailed balance is broken, this reversiblity is broken.  In this 
case, the path from starting to destination point and then back again to the starting point forms a loop, 
and the sense of rotation around the loop is determined by the vorticity \cite{Dannenberg_PRL_2014, 
Ghanta_PRE_2017}. 


\section{ The Geometric Action }
\label{sec: 2}

We begin by reviewing the statistics of noise driven trajectories, in the spirit that was formulated many 
years ago by Feynman and Hibbs \cite{Feynman_1965}.  The state space is ${\mathbb{R}}^N$, and 
trajectories are curves $\mathbf{x} = \mathbf{x}(t)$ in state space parametrized by time $t$.  The 
$\mathbf{x}(t)$ satisfy a Langevin equation
\begin{equation}
\dot{\mathbf{x}} - \mathbf{u}(\mathbf{x}) = \sqrt{\epsilon}\sigma\mathbf{w}(t).
\end{equation}
Here, $\mathbf{u}(\mathbf{x})$ is a given flow vector field on state space, the components of 
$\mathbf{w}(t)$ are independent unit white noises, and $\sigma$ is a noise tensor, assumed uniform 
and constant.  $\epsilon > 0$ is a gauge parameter, so we can formalize the small noise limit $\epsilon > 
0$.  The Langevin equation (1) with uniform and constant noise tensor $\sigma$ is physically 
appropriate when the noise represents external forcing as a function of time, by degrees of freedom not 
included in $\mathbf{x}$.

The trajectories we consider are solutions of (1) which pass through two given points $\mathbf{a}$ and 
$\mathbf{b}$ of state space in succession.  There is no restriction on the time of flight $T$ from 
$\mathbf{a}$ to $\mathbf{b}$.  Due to autonomy, we can set the origin of time so $\mathbf{x}(0) = 
\mathbf{a}$.  Then $\mathbf{x}(T) = \mathbf{b}$.  
The essential idea is that the relative probability of such a trajectory from $\mathbf{a}$ to $\mathbf{b}$ 
is the same as the relative probability of the white
noise $\mathbf{w} (t)$ in $0 < t < T$ which produces it. The relative probability of a noise sequence 
$\mathbf{w} (t)$ in $0 < t< T$ is expressed formally as
\begin{equation}
    e^{-\frac{1}{4} \int_0^T | \mathbf{w} (t) | ^ 2 dt}.
\end{equation}
For the intuiton behind (2), model the sample space of white noises in $0 < t < T$ by piecewise constant 
functions.  Any one component of $\mathbf{w} (t)$ is represented by $w(t) \equiv w_j$ in $(j - 1) \Delta 
t < t < j \Delta t$.  Here, the time increment is $\Delta t = \frac{T}{N}$ for some positive integer $N$.  
There are $N$ time intervals of
piecewise constant $w(t)$ in $0 < t < T$.  The constant values of $w(t)$ in different time intervals are 
statistically independent, and the probability density of each $w_j$ is proportional to the Gaussian
\begin{equation}
e^{- \frac{{w_j}^2}{2(\frac{2}{ \Delta t})}} = e^{ - \frac{1}{4} {w_j}^2 \Delta t},             
\end{equation}
with mean square $\frac{2}{\Delta t}$.  Due to the independence of the $w_j$ for different $j$, the 
probability density in the ${\mathbb{R}}^N$ of $N$-tuples $(w_1, ...w_N)$
which characterize whole noise sequences in $0 < t < T$ is the product of the Gaussians (3),
\begin{equation}
    e^{ - \frac{1}{4}( \sum_1^N {w_j}^2 ) \Delta t} .
\end{equation}
We now see that the formal integral (2) is a shorthand reminder of this construction.  The choice of 
mean square $\frac{2}{ \Delta t}$ for each $w_j$ gives the correct behavior of the Brownian motion,
\begin{equation}
B(t) := \int_0^t w(t') dt'.
\end{equation}
At $t = n \Delta t$, we have
\begin{equation}
\langle B^2 (t) \rangle = n (\frac{2}{\Delta t}) (\Delta t)^2 = 2n \Delta t = 2t.                                                                  
\end{equation}
In this little review, we have presented (2) as the relative probability of noise sequences subject to no 
constraints.  The noise sequences $\mathbf{w} (t)$ which actually produce a trajectory with starting 
point $\mathbf{a}$ and endpoint $\mathbf{b}$ are in a restricted class - i.e., they must obey the 
boundary conditions, and we assume that the relative probabilities within this restricted class are still 
characterized by (2).

The relative probability (2) is converted into a functional of the trajectory $\mathbf{x} = \mathbf{x} (t)$ 
in $0 < t < T$ simply by ``solving" the stochastic differential equation (1) for $\mathbf{w} (t)$,
\begin{equation}
    \mathbf{w} (t) = \frac{1}{\sqrt{\epsilon}} {\sigma}^{-1} (\dot{\mathbf{x}} - \mathbf{u} ),
\end{equation}
and substituting this $\mathbf{w} (t)$ into (2).  The relative probability is thereby expressed as
\begin{equation}
    e^{ - \frac{1}{\epsilon} S [ \mathbf{x} (t) ]} ,
\end{equation}
where $S[ \mathbf{x} (t) ]$ is the \textit{stochastic action functional}, given by
\begin{equation}
    S[ \mathbf{x} (t) ] := \frac{1}{4} \int_0^T (\dot{\mathbf{x}} - \mathbf{u}) \cdot D^{-1} 
(\dot{\mathbf{x}} - \mathbf{u} ) dt.
\end{equation}
Here,
\begin{equation}
    D := \sigma {\sigma}^T
\end{equation}
is the \textit{diffusion tensor} associated with the noise tensor $\sigma$.  In this whole train of thought, 
we are assuming that $\sigma$ is invertible \cite{Heymann_2015}.

The global minimizer $\mathbf{x}(t)$ of the action (9) is called the \textit{most probable trajectory}.  
The minimization of the action determines the \textit{geometric curve} $C$ that $\mathbf{x}(t)$ traces 
out the path from $\mathbf{a}$ to $\mathbf{b}$, and also the time sequence of positions along $C$.  In 
particular, the flight time $T$ is precisely determined by (21) below.  The traditional analysis 
\cite{Luchinsky_RPP_1998} begins with the Euler-Lagange equation of the action (9), with $T$ fixed.  The 
Lagangian is
\begin{equation}
L(\mathbf{x}, \mathbf{v} := \dot{\mathbf{x}}) := \frac{1}{4}(\mathbf{v} - \mathbf{u})\cdot D^{-1} 
(\mathbf{v} - \mathbf{u}).
\end{equation}
Due to automomy, there is a conserved ``energy"
\begin{equation}
h(\mathbf{x}, \mathbf{v})  = \mathbf{v} \cdot \mathbf{\nabla}_{\mathbf{v}} L - L = 
\frac{1}{4}(\mathbf{v} \cdot D^{-1} \mathbf{v} - \mathbf{u} \cdot D^{-1} \mathbf{u}).
\end{equation}
Some presentations carry out a Legendre transformation from Lagrangian to Hamiltonian dynamics 
\cite{Dykman_JCP_1994, Luchinsky_RPP_1998}.  Here, we stay within the Lagrangian framework.  Think 
of $D^{-1}$ as the metric tensor of the ${\mathbb{R}}^N$ in which $\dot{\mathbf{x}}$ and 
$\mathbf{u}$ live.  The associated inner product is
\begin{equation}
\mathbf{f} \bullet \mathbf{g} := D^{-1}  \mathbf{f} \cdot \mathbf{g},
\end{equation}
and the squared ``length" of $\mathbf{f}$ is
\begin{equation} 
| \mathbf{f} |^2 := \mathbf{f} \bullet \mathbf{f}.
\end{equation}
In geometric notation, the action (9) is
\begin{eqnarray}
S &=& \frac{1}{4} \int_{0}^{T} | \dot{\mathbf{x}} - \mathbf{u} |^2 dt  \nonumber \\ &=& \frac{1}{4} 
\int_{0}^{T} (\dot{\mathbf{x}} \bullet \dot{\mathbf{x}} - 2\mathbf{u} \bullet \dot{\mathbf{x}} + 
\mathbf{u} \bullet \mathbf{u}) dt \nonumber \\ &=& \frac{1}{4} \int_{0}^{T} (| \dot{\mathbf{x}} |^2 - 2| 
\dot{\mathbf{x}} | |\mathbf{u} | + | \mathbf{u} |^2) dt  +  \frac{1}{2} \int_{0}^{T} (| \mathbf{u} | | 
\dot{\mathbf{x}} | - \mathbf{u} \bullet \dot{\mathbf{x}}) dt, \nonumber
\end{eqnarray}
or, finally,
\begin{equation}
S =  \frac{1}{4} \int_{0}^{T} (| \dot{\mathbf{x}} | - | \mathbf{u} | )^2 dt +                                                                                                                
\frac{1}{2} \int_{0}^{T} ( | \mathbf{u} | | \dot{\mathbf{x}} | - \mathbf{u} \bullet \dot{\mathbf{x}} ) dt.
\end{equation}
The first integral on the right hand side achieves its minimum value of \textit{zero} if the speed $| 
\dot{\mathbf{x}}(t) |$ of the trajectory matches the speed $|\mathbf{u}(\mathbf{x}(t))|$ of the 
deterministic flow,
\begin{equation}
| \dot{\mathbf{x}}(t)| = | \mathbf{u}(\mathbf{x}(t)) |.
\end{equation}
Given (16), the action (15) reduces to
\begin{equation}
S = \frac{1}{2} \int_{0}^{T} ( | \mathbf{u} | | \dot{\mathbf{x}} | - \mathbf{u} \bullet \dot{\mathbf{x}} ) 
dt.
\end{equation}
In (17), we recognize
\begin{equation}
\mathbf{dx} = \dot{\mathbf{x}} dt,
\end{equation}
and 
\begin{equation}
ds = | \mathbf{dx} | = | \dot{\mathbf{x}} | dt.
\end{equation}
as increments of displacement and arclength in elapsed time $dt$, respectively.  
Hence, the action (9) reduces to a geometric line integral along the curve $C$ connecting $\mathbf{a}$ 
to $\mathbf{b}$,
\begin{equation}
S = \frac{1}{2} \int_C (| \mathbf{u} | ds - \mathbf{u} \bullet \mathbf{dx} ).
\end{equation}
In summary, the most probable trajectory $\mathbf{x}(t)$ from $\mathbf{a}$ to $\mathbf{b}$ proceeds 
with deterministic speed as in (16), and the geometric curve $C$ traced out by $\mathbf{x}(t)$ 
minimizes the line integral (20).

We acknowledge some collateral insights:  since the speed along $C$ is the deterministic speed,
the time of flight from $\mathbf{a}$ to $\mathbf{b}$ is
\begin{equation}
T = \int_C \frac{ds}{| \mathbf{u}|}.
\end{equation}
A further consequence of (16) is that the most probable trajectory lives on the null surface of the energy 
(12) in $\mathbf{x}, \dot{\mathbf{x}}$ space.  In geometric notation, (12) reads
\begin{equation}
h = \frac{1}{4}(|\dot{\mathbf{x}}|^2 - |\mathbf{u}|^2)
\end{equation}
and (16) implies $h = 0$.  In the important special case where the most probable paths connect 
\textit{critical points}, we can deduce directly that $h \equiv 0$ and hence
$| \dot{\mathbf{x}}| = | \mathbf{u} |$ because $\dot{\mathbf{x}}$ and $\mathbf{u}$ both vanish at 
critical points.  But the null Hamiltonian character of most probable paths is more general.  In the 
preceeding argument, we have seen that the speed along the least action flucatuational path between 
\textit{any} two endpoints equals the deterministic speed. Hence the null Hamiltonian character of most 
probable paths applies to any pair of endpoints, be they critical or not.

The propagation along $C$ at deterministic speed seems deeply peculiar when we reflect that it is an 
asymptotic result of the small noise limit $\epsilon \rightarrow 0.$
Notice that we can halve the noise amplitude, and half the noise still drives the \textit{same}
speed of propogation along $C$.  Here is another collateral insight:  A change in $\epsilon$ can be 
absorbed by scaling the diffusion tensor $D$.
Look at the contribution to the time of flight $T$ from a small segment $\mathbf{dx}$ of $C$.  The 
increment of arclength is
\begin{equation}
ds = \sqrt{D^{-1} \dot{\mathbf{x}} \cdot \dot{\mathbf{x}}} dt,  
\nonumber
\end{equation}
the speed is
\begin{equation}
| \mathbf{u} | = \sqrt{D^{-1} \mathbf{u} \cdot \mathbf{u}},
\nonumber
\end{equation}
and it is clear that the scaling of $D$ does not change the time increment $dt = \frac{ds}{| \mathbf{u} 
|}.$

A qualitative explanation might go like this:  The noise induced sequence of kicks that drives the speed 
$| \mathbf{u} |$ must have some ``winning" combination of strength and unidirectionality, and it must 
last throughout the flight time $T$.  As the noise amplitude goes to zero, the time duration of a 
``winning streak"
($\mathbf{x}(t)$ goes from $\mathbf{a}$ to $\mathbf{b} )$ remains near $T$, cf. (21), but the expected 
time \textit{between} winning streaks becomes exponentially large.


\section{Broken detailed balance and the geometry of least action paths}

We begin with some details behind the overview of detailed balance in the introduction.  Let $\rho 
(\mathbf{x}, t)$ be the ensemble probability density.  The probability current generated by the 
stochastic dynamics (1) is
\begin{equation}
\mathbf{J} := \rho \mathbf{u} - D \mathbf{\nabla} \rho .
\end{equation}
Here, $\mathbf{u} ( \mathbf{x} )$ is the velocity field of the stochastic differential equation (1), and $D$ 
is the diffusion tensor (10).  Take a bounded region $\mathcal{R}$ of state space, and impose the 
impermeable boundary condition, $\mathbf{J} \cdot \mathbf{n} = 0$ on $\partial \mathcal{R}$.
Then the probability density in $\mathcal{R}$ asymptotes to a time independent steady state $\rho = 
\rho ( \mathbf{x})$.  In a detailed balance system, the probability current is not only divergence-free, 
but vanishes identically, so
\begin{equation}
\rho \mathbf{u} = D \mathbf{\nabla} \rho
\end{equation}
throughout $\mathcal{R}$.  It follows from (24) that
\begin{equation}
\mathbf{\nabla} ( \log \rho ) = D^{-1} \mathbf{u}
\end{equation}
or, equivalently, the \textit{stochastic vorticity tensor} $\Omega$, with components
\begin{equation}
{\Omega}_{ij} := \partial_j (D^{-1} \mathbf{u} )_i - \partial_i (D^{-1} \mathbf{u} )_j,
\end{equation}
vanishes. 

To see the role of vorticity in the geometry of most probable paths, we write down the variational 
differential equation of the geometric action (20).  Let $\mathbf{x} = \mathbf{x} (s)$ be the parametric 
representation of a most probable path with respect to the arclength $s$.  Then, the variational 
equation is
\begin{equation}
\frac{d}{ds} (|\mathbf{u}| D^{-1} \frac{\mathbf{dx}}{ds} ) - \nabla | \mathbf{u} | = \Omega 
\frac{\mathbf{dx}}{ds}.
\end{equation}
Evoking the close analogy to geometric optics \cite{Born_1980}, any curve traced out by a solution of 
(27) is called a \textit{ray}.  The LHS of (27) is invariant under orientation reversal $(s \rightarrow -s)$, 
but the RHS changes sign if the vorticity is nonzero.  Hence, the forward and backward paths between 
two endpoints in a region of nonzero vorticity are generally not reversals of each other.  The addition of 
forward and backward paths makes a closed loop, whose orientation is determined by the vorticity 
tensor. 

A simple local analysis of rays informs the structure of forward plus backward loops whose endpoints 
are close to each other.  Let $C: \mathbf {x} = \mathbf{x} (s)$ be a curve parametrized by arclength $s$.  
Set the origin of arclength so as to mark a given point $\mathbf{X}$ on $C$.  In the Taylor series of 
$\mathbf{x} (s)$ as $s \rightarrow 0$, we have
\begin{equation}
    \mathbf{x} (s) = \mathbf{X} + \mathbf{x} ' (0) s + \frac{1}{2} \mathbf{x} '' (0) s^2 + O( s^3),
\end{equation}
where the tangent $\mathbf{x} ' (0)$ at $\mathbf{X}$ is a unit vector, $| \mathbf{x} ' (0) | = 1$ (since 
the expansion is with respect to arc length), and the second derivative is orthogonal to the tangent, 
$\mathbf{x} ' (0) \bullet \mathbf{x} '' (0) = 0$.  If $\mathbf{x} '' (0) \ne 0$, we see that $C$ near 
$\mathbf{X}$ is asymptotic to a parabola in the plane spanned by $\mathbf{x}'(0)$ and $\mathbf{x} '' 
(0)$. Now look at two particular rays $C_+ : {\mathbf{x}}_+ (s)$ and $C_- : {\mathbf{x}}_- (s)$ which 
meet at $\mathbf{X}$ at $s = 0$, with opposite and equal tangents, $\mathbf{\hat{t}} := {\mathbf{x}}_+ ' 
(0) = - {\mathbf{x}}_- ' (0)$. From the ray equation (27), it follows that the vorticity induces a 
\textit{difference} of second derivitives,
\begin{equation}
[ \mathbf{x} '' (0) ] := {\mathbf{x}}_+''(0) - {\mathbf{x}}_-''(0) = \frac{2}{| \mathbf{u} |} D \Omega 
\mathbf{\hat{t}}.
\end{equation}
To see this, we evaluate (27) at $s = 0$ to obtain $| \mathbf{u} | D^{-1} \mathbf{x}''(0) + (\mathbf{x}'(0) 
\cdot \nabla)|\mathbf{u}|D^{-1} \mathbf{x}'(0) - \nabla | \mathbf{u} | = \Omega \mathbf{x}'(0)$.  Note 
that the second and third terms on the LHS of this equation are invariant under reversal, $(s \rightarrow 
-s)$.  Thus, taking differences evaluated along $C_+$ and $C_-$,  we can write $| \mathbf{u} | D^{-1} 
[\mathbf{x}''(0)] = \Omega [\mathbf{x}'(0)] = 2 \Omega \mathbf{\hat{t}}$, from which (29) follows 
directly.

\begin{figure}[]
\includegraphics[scale=0.7]{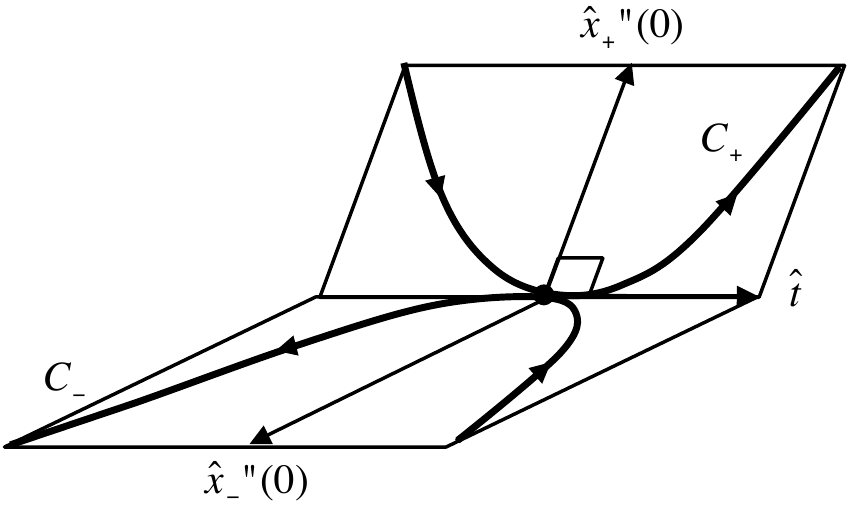}
%
%
\caption{Illustration of trajectories $C+$ and $C_-$ which just touch at position $\mathbf{X}$ with 
oppositely directed tangent vectors.}
\label{fig:1}       
\end{figure}

\begin{figure}
\includegraphics[scale=0.8]{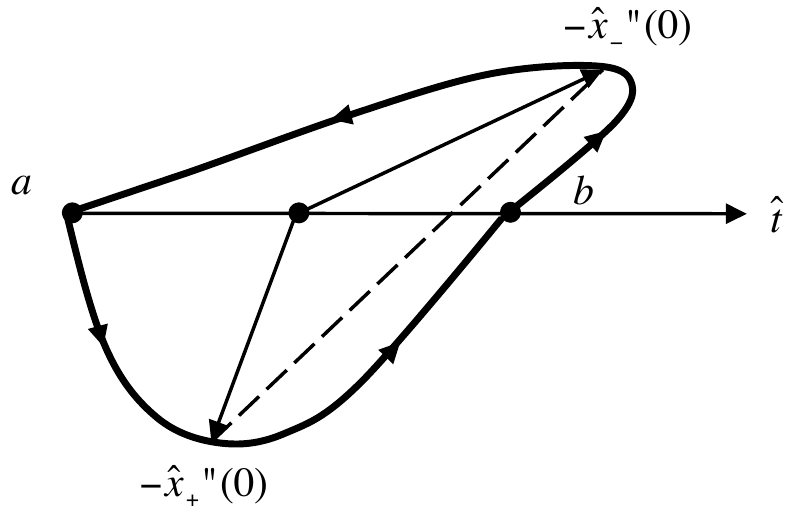}
%
%
\caption{Illustration of ``forward" and ``backward" loops for endpoints $\mathbf{a}$ and $\mathbf{b}$ 
in the neighborhood of $\mathbf{X}$, obtained by translating $C_+$ in the $-\mathbf{x}_{+}''(0)$ 
direction and $C_-$ in the $-\mathbf{x}_{-}''(0)$ direction.}
\label{fig:2}       
\end{figure}

The antisymmetry of $\Omega$ implies that $[ \mathbf{x} '' (0) ]$ is orthogonal to $\mathbf{\hat{t}}$, 
as it must be, since ${\mathbf{x}}_+''(0)$ and ${\mathbf{x}}_-''(0)$ are.  For $\Omega \mathbf{\hat{t}} 
\ne 0$, we can visualize $C_+$ and $C_-$ near $\mathbf{X}$ as two parabolas tangent to each other at 
$\mathbf{X}$, but bending in different planes, $C_+$ in the plane of $\mathbf{\hat{t}}$ and 
${\mathbf{x}}_+''(0)$, and $C_-$ in the plane of
 $\mathbf{\hat{t}}$ and ${\mathbf{x}}_-''(0)$ as shown in Figure 1.  Now take two endpoints 
$\mathbf{a}$ and $\mathbf{b}$ close to $\mathbf{X}$, and take $\mathbf{\hat{t}}$ to be parallel to 
$\mathbf{b} - \mathbf{a}$.  By small translations of the parabolas in Figure 1 in the ${-
\mathbf{x}}_+''(0)$ and ${-\mathbf{x}}_-''(0)$ directions, we can asymptotically construct the forward 
plus backward loop with these endpoints.  Pictorially, the difference $\mathbf{x}_+''(0) - \mathbf{x}_-
''(0)$ corresponds to the ``openness" of the constructed fluctuation loop as shown in Figure 2.  The 
dashed vector marks the displacement between the turning points of the $\mathbf{a}$ to $\mathbf{b}$ 
and $\mathbf{b}$ to $\mathbf{a}$ parabolas; it is proportional to $[ \mathbf{x} '' (0) ] = \frac{2}{| 
\mathbf{u} |} D \Omega \mathbf{\hat{t}}$.
 
 In two dimensions, we can develop further intuition.  In that case, the vorticity tensor takes the form 
 \begin{equation}
 \Omega = 
 \left[
 \begin{array}{cc}
 0 & -\omega\\
 \omega & 0
 \end{array}
 \right],
 \end{equation}
 where $\omega := \partial_1 (D^{-1} \mathbf{u} )_2 - \partial_2(D^{-1} \mathbf{u})_1$, is the 
\textit{scalar vorticity}.  For $\omega > 0$, (27) is the generator of a counterclockwise rotation.  Figure 3 
is the two dimensional version of Figure 2, drawn assuming scalar diffusion ($D$ proportional to the 
identity) and $\omega > 0$.  The counterclockwise circulation around the loop agrees with the 
counterclockwise rotation associated with the vorticity.  This is the choice you would make to reduce 
``headwinds" and thereby reduce the stochastic action.
 
\begin{figure}[]
\includegraphics[scale=0.8]{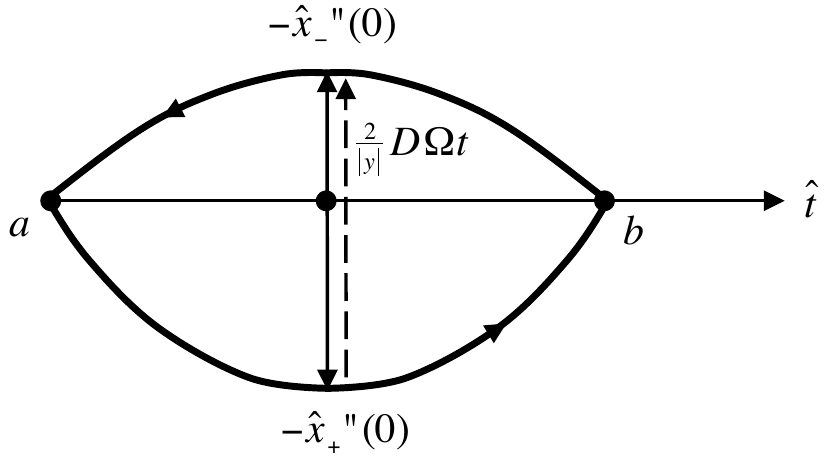}
\caption{Depiction of forward and backward loops in the two-dimensional case.}
\label{fig:3}       
\end{figure}


\section{Vorticity-Induced Bifurcations}
\label{sec:4}

The global effects of vorticity-induced bending of rays are most simply displayed in a class of two-
dimensional examples we call \textit{pure shear}:  we have isotropic diffusion $D = I$, and the flows 
$\mathbf{u}$ on $\mathbb{R}^2$ are\begin{equation}
    \mathbf{u} = u(y) \hat{\mathbf{y}},
\end{equation}
where the \textit{velocity profile} $u(y)$ is a given positive function.  The $x$-component of the ray ODE 
(27) is
\begin{equation}
    \frac{d}{ds} ( u \frac{dx}{ds} ) = u'(y) \frac{dy}{ds},
\end{equation}
which has the first integral
\begin{equation}
    u(y) (1 - \frac{dx}{ds} ) = \sigma = constant.
\end{equation}
(33) is analogous to Snell's law in optics: that is, $ \frac{dx}{ds} = \cos \theta$, where $\theta$ is the 
local angle of the ray with respect to $ \hat{\mathbf{x}}$. 
Hence, (33) implies that $\theta$ changes with elevation $y$ due to gradients of the
velocity profile $u(y)$.  In this sense, we have ``refraction by shear."  Due to the first integral (33), the 
geometric action of a ray $C$ has the simple expression
\begin{equation}
    S = \int_C u ds - u dx = \int_C u(1 - \frac{dx}{ds} ) ds = \sigma L,
\end{equation}
where $L$ is the arclength of $C$.  Deterministic paths are immediately recognized from the effective 
Snell's law (33) and (34):  The geometric action vanishes if the constant $\sigma$ is zero, and then $ 
\frac{dx}{ds} \equiv 1$, which corresponds to horizontal straight lines oriented to the right.  These lines 
are the obvious integral curves of the shear flow.

Next, we examine rays that undergo a net displacement in the negative x-direction, which is ``against 
the wind."  By the mean value theorem, such a ray must have a \textit{turning point} ${\mathbf{x}}_*$, 
where $\frac{dx}{ds} = -1.$  Evaluating
(33) at the turning point, we have $\sigma = 2 u_* := 2 u(y_*)$, where $y_*$ is the elevation of the 
turning point.  Hence, (33) can be written as
\begin{equation}
u \frac{dx}{ds} = u - 2 u_*,
\end{equation}
and the geometric action becomes
\begin{equation}
S = 2 u_* L.
\end{equation}
We can determine a governing equation for the vertical component $y(s)$ of $\mathbf{x} (s)$, by 
evoking the geometric identity $ ( \frac{dx}{ds} )^2 + ( \frac{dy}{ds} )^2 = 1$, and using the effective 
Snell's law (35):
\begin{equation}
u^2 ( \frac{dy}{ds} )^2 = 4 u_* (u(y) - u_*).
\end{equation}
Since $u(y)$ is the deterministic speed along the ray, we recognize $u \frac{d}{ds}$ as the time derivitive 
$ \frac{d}{dt}$, and we write (35) and (37) respectively as
\begin{equation}
\dot{x} = u(y) - 2 u_*,
\end{equation}
\begin{equation}
{\dot{y}}^2 = 4 u_* (u(y) - u_*).
\end{equation}
The second equation is the first integral of the second order differential equation
\begin{equation}
\ddot{y} = 4 u_* u'(y).
\end{equation}
Here, $-u'(y)$ is the scalar vorticity of the shear flow, so (40) directly expresses "bending by vorticity."  

We can construct rays explicitly for the instructive case of quadratic shear flow, with
\begin{equation}
u(y) = \frac{1}{2} (y^2 + {\epsilon}^2 ),
\end{equation}
where $\epsilon$ is a positive parameter.  The minimum positive $x$-velocity happens along the $x$-
axis.  We expect that rays with a start to finish displacement in the negative $x$-direction bend toward 
the $x$-axis, to take advantage of the reduced headwind there.  Such a ray must have a turning point.  
By the $x$ and $t$ translation invariance of (38) and (39), we can use the turning point to mark the 
origins of $x$ and $t$, so we have initial conditions $x(0) = 0, y(0) = y_*, \dot{y} (0) = 0$.  The solutions 
of (38), 39) with these initial conditions are
\begin{equation}
y = y(t, y_*) = y_* \cosh \tau,
\end{equation}
\begin{equation}
x = x(t, y_*) = \frac{{y_*}^2}{4 \sqrt{{y_*}^2 + \epsilon^2}} (\cosh \tau \sinh \tau - \tau) - 
\frac{\sqrt{{y_*}^2 + \epsilon^2}}{2} \tau.
\end{equation}
Here, $\tau$ is the scaled time
\begin{equation}
\tau := \sqrt{{y_*}^2 + \epsilon^2} t.
\end{equation}
For each value of $y_*$ we obtain a corresponding ray represented by a parametric curve
\begin{equation}
\mathbf{x} (t) = x(t, y_*) \hat{\mathbf{x}} + y(t, y_*) \hat{\mathbf{y}}.
\end{equation}

\begin{figure}[t]
\includegraphics[scale=0.5]{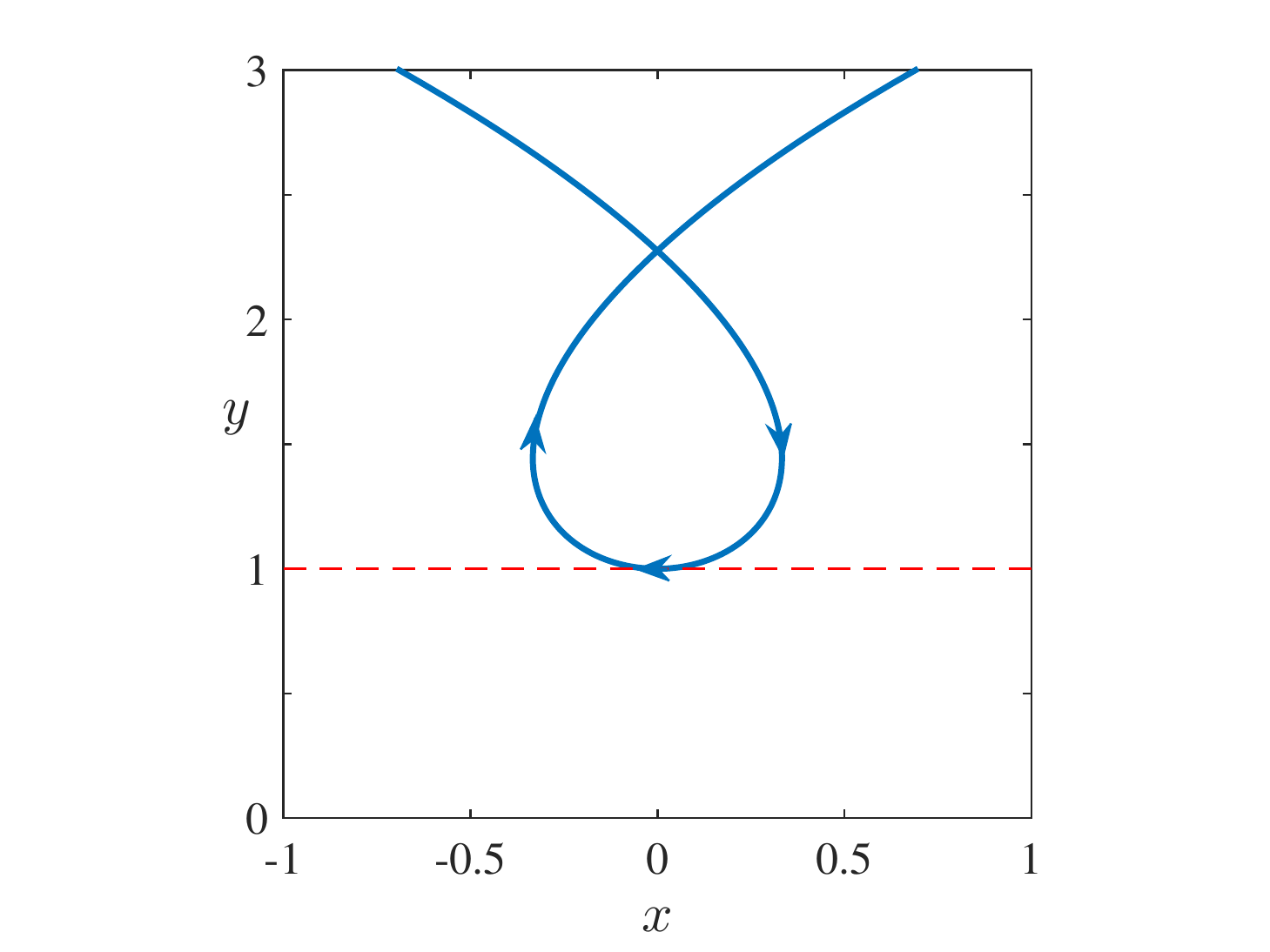}
\caption{Transition path ray for the case $y_* = 1$ and $\epsilon = 0.3$.}
\label{fig:4}       
\end{figure}

We present a pictorial narrative of these rays for the case $\epsilon = 0.3$, and this shows how 
segments of rays produce net displacements in the negative $x$-direction.
Figure 4 shows the ray with turning point elevation $y_* = 1$.  The most striking feature is the teardrop-
shaped loop with counterclockwise orientation.  The ``teardrop" persists for all $y_* > 0$.  From (38), we 
see that the vertical tangents happen at the elevation $y$ for which the shear velocity $u(y)$ is twice as 
high as at the elevation of the turning point.  As $y_* \rightarrow 0$, the bottom of the teardrop 
between the vertical tangents flattens out into a long, left-oriented segment close to the $x$-axis.  We 
see this in Figure 5, which depicts the ray with $y_* = .01$. 

\begin{figure}
\includegraphics[scale=0.5]{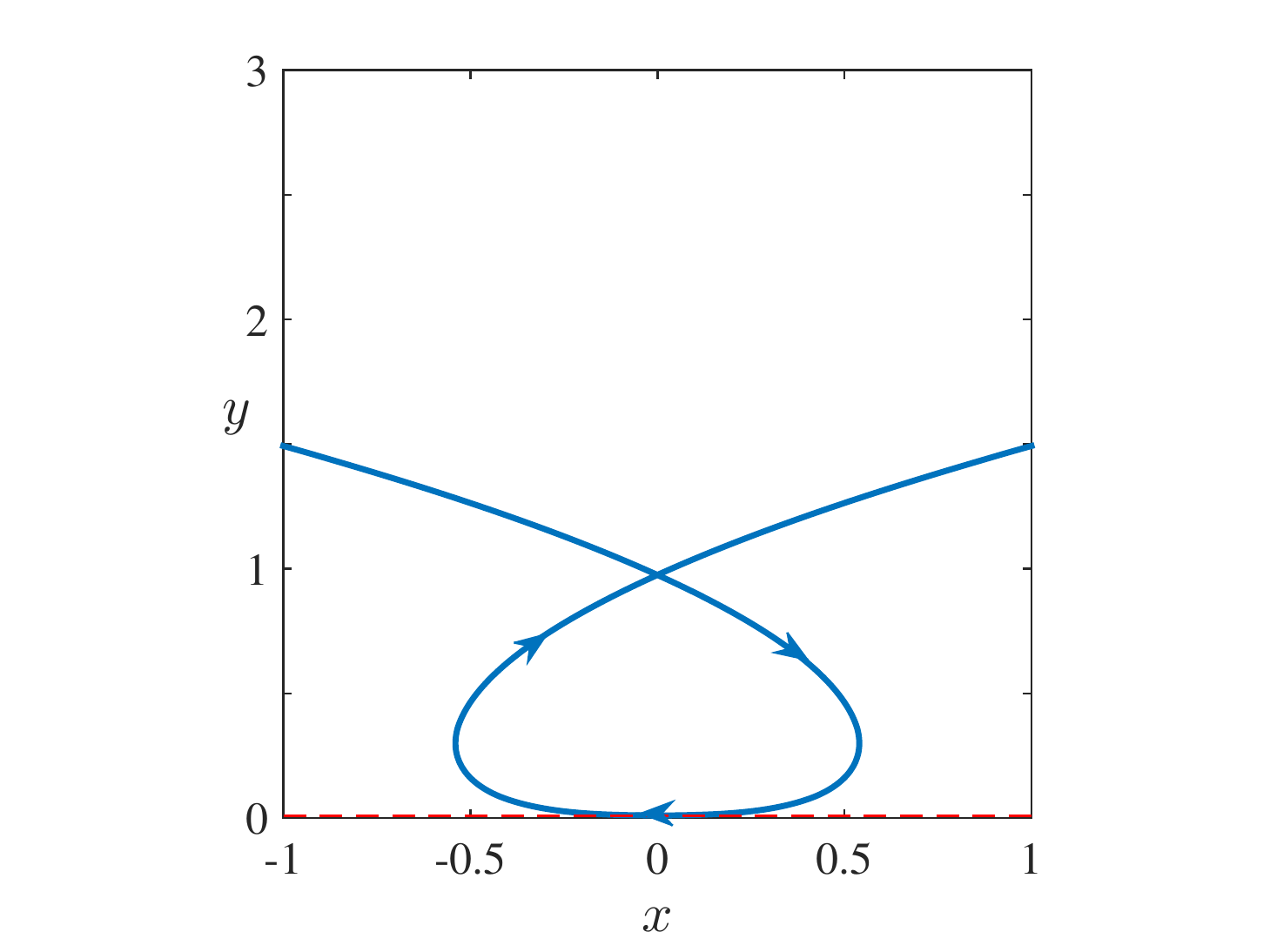}
\caption{Transition path ray for the case $y_* = 0.01$ and $\epsilon = 0.3$.}
\label{fig:5}       
\end{figure}

A ray which interpolates between given starting and ending points $\mathbf{a}$ and $\mathbf{b}$ is 
called a \textit{connector}.  We examine the connectors with
\begin{equation}
\mathbf{a} = \frac{l}{2} \hat{\mathbf{x}} + \hat{\mathbf{y}} ,\mathbf{b} = - \frac{l}{2} \hat{\mathbf{x}} + 
\hat{\mathbf{y}},
\end{equation}
which makes a net displacement of length $l$ in the negative $x$-direction. These connectors are 
segments of the ``teardrop," symmetric about the $x$-axis.  Given $l$, there are discrete choices for the 
turning point elevations $y_*$ of possible connectors.
Let $\tau = \tau_b > 0$ be the value of scaled time marking the ending point $\mathbf{b}$.  By 
symmetry, $\tau = -\tau_b$ marks the starting point $\mathbf{a}$, and the connector passes through 
the turning point $y_* \hat{\mathbf{y}}$ at $\tau = 0$.  At $\tau = \tau_b$, the elevation $y$ is unity, so 
(42) implies
\begin{equation}
    y_* \cosh \tau_b = 1.
\end{equation}
 Given $0 < y_* < 1$, we solve for $\tau_b$,
 \begin{equation}
     \tau_b = \log { \frac{1 + \sqrt{1 - y_*^2}}{y_*}} .
 \end{equation}
 The actual time of flight from $\mathbf{a}$ to $\mathbf{b}$ is
 \begin{equation}
     T = \frac{2\tau_b}{\sqrt{y_*^2 + \epsilon^2}} + \frac{2}{\sqrt{y_*^2 + \epsilon^2}}\log { \frac{1 + 
\sqrt{1 - y_*^2}}{y_*}}
 \end{equation}
 At $\tau = -\tau_b$, $x = \frac{l}{2}$, and then (43) implies
 \begin{equation}
     \frac{l}{2} = \frac{y_*^2}{4 \sqrt{y_*^2 + \epsilon^2}} (\tau_b - \cosh \tau_b \sinh \tau_b) + 
\frac{\sqrt{y_*^2 + \epsilon^2}}{2} \tau_b.
 \end{equation}
 Substituting for $\tau_b$ from (48), this reduces to
 \begin{equation}
 \begin{split}
     l = (\sqrt{y_*^2 + \epsilon^2} +  \frac{y_*^2}{2 \sqrt{y_*^2 + \epsilon^2}})\log  \Big[ \frac{1 + \sqrt{1 - 
y_*^2}}{y_*} \Big] \\ - \frac{\sqrt{1 - y_*^2}}{2\sqrt{y_*^2 + \epsilon^2}} .
     \end{split}
 \end{equation}
 The graph of $y_*$ vs $l$ in Figure 6 based on (51) is plotted with $\epsilon = 0.3$, but is qualitatively 
correct for $0 < \epsilon << 1$.  For connectors with a net displacement in the negative $x$-direction, 
the relevant portion of this graph has $l > 0$, marked by the solid curves.   There are three branches of 
roots for $y_*$ as functions of $l$, which we've labeled i, ii, and iii.  Branches i and ii coelesce in a 
saddle-node bifurcation as $l \rightarrow l_c \simeq 0.51$.  Furthermore, the branches i and ii have 
definite limits as $\epsilon \rightarrow 0$, and the saddle-node bifurcation survives with $l_c 
\rightarrow 0.4852$.  Figure 7 depicts the connecting pathways corresponding to branches i, ii and iii for 
$\epsilon = 0.3$.  
 
\begin{figure}
\includegraphics[scale=0.5]{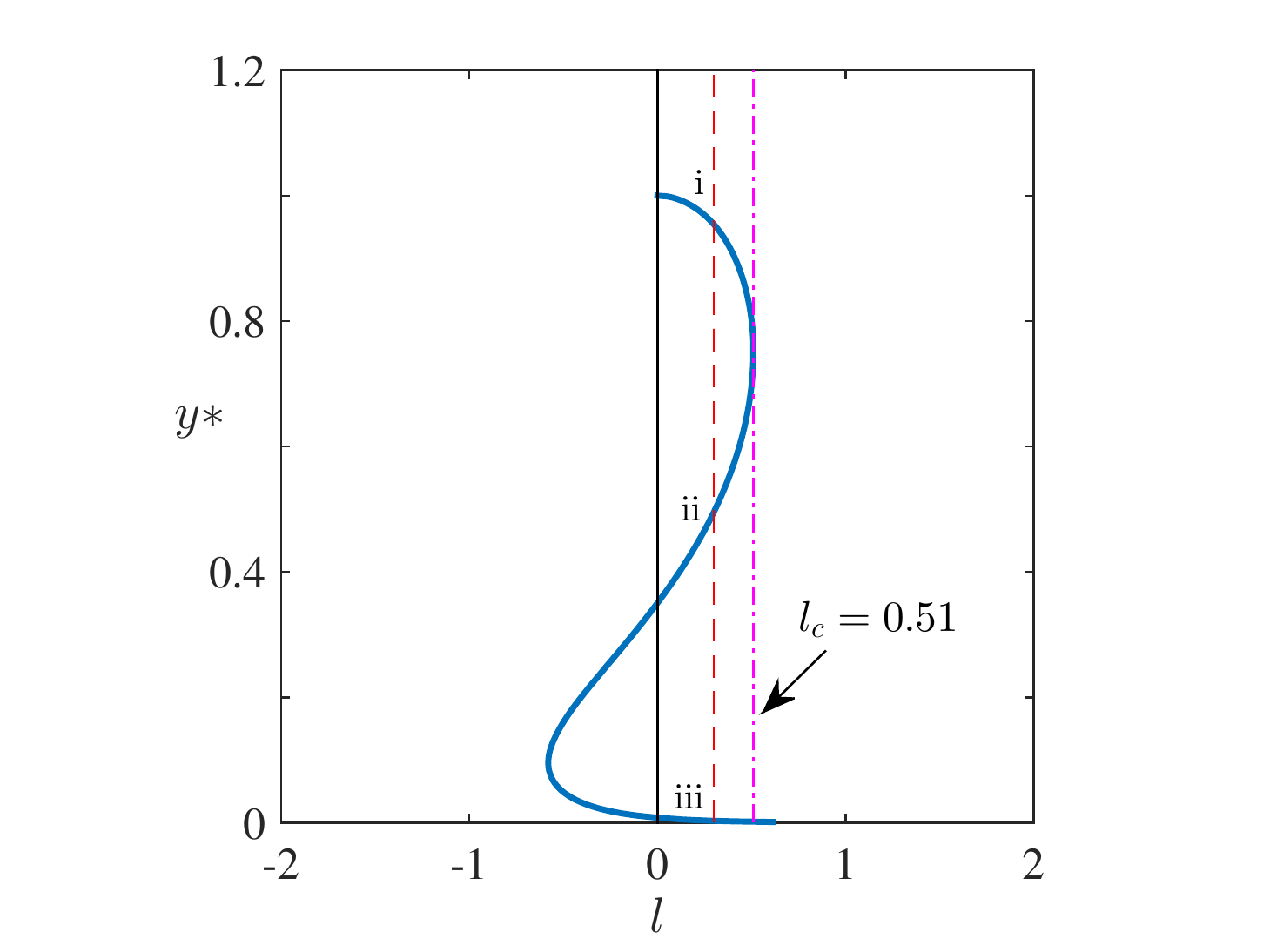}
\caption{Plot of Equation (53) showing the allowed values of $y_*$ as a function of $l$ and $\epsilon = 
0.3$. The turning point at $l_c$ corresponds to a saddle-node bifurcation in which two transition paths 
with endpoints $\mathbf{a}$ and $\mathbf{b}$ collide and annihilate.}
\label{fig:6}       
\end{figure}

 Which connector has the smallest geometric action?  We can compute the arclength $L$ of connectors, 
and then their geometric actions from $S = 2 u_* L$ as in (36).  Since $u = u(y)$ is the speed along a ray, 
the arclength of a connector $C$ is
 \begin{equation}
     L = \int_C u dt = \int_C (\dot{x} + 2 u_*) dt = -l + 2 u_* T.
 \end{equation}
 Substituting into (42) the time of flight $T$ from (49), and $l$ from (51), we obtain for the action
 \begin{equation}
     S = \frac{\sqrt{y_*^2 + \epsilon^2}}{2} ((y_*^2 + 2 \epsilon^2)\log { \frac{1 + \sqrt{1 - y_*^2}}{y_*}} + 
\sqrt{1 - y_*^2} ) .
 \end{equation}
 From (53), we find that the shorter branch i connector has the smallest geometric action.  Apparently, 
the branch ii  connector is a stationary point of geometric action, but not a minimum.

\begin{figure}
\includegraphics[scale=0.5]{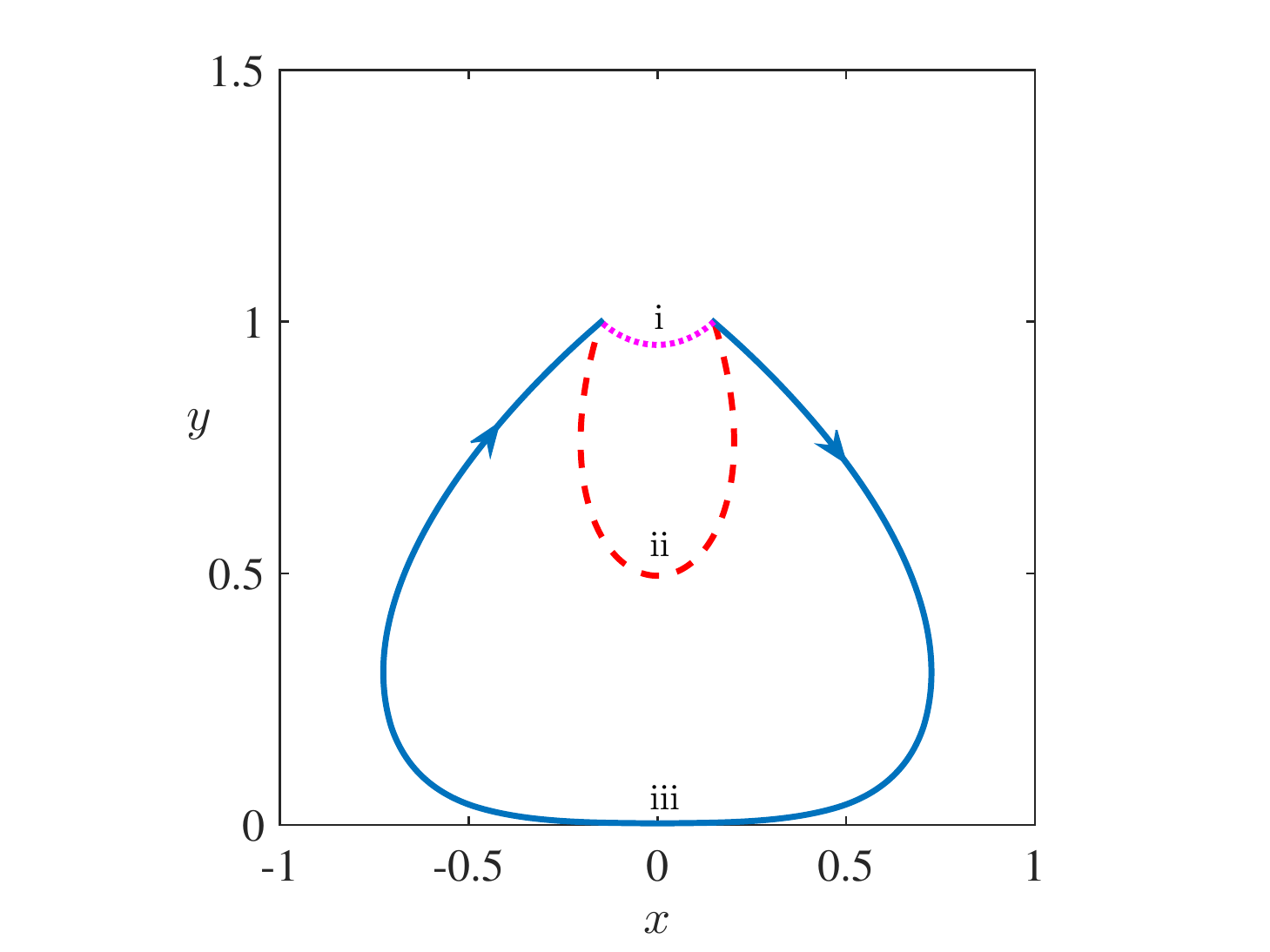}
\caption{Plot of the three connector solutions corresponding to the three branches of Figure 6 for $l = 
0.3$ and $\epsilon = 0.3$.}
\label{fig:7}       
\end{figure}

Branch iii in Figure 7 and its corresponding connector persist for all $l > 0$. In the limit $\epsilon 
\rightarrow 0$, branch iii of the $y_*$ vs $l$ relation is asymptotic to
\begin{equation}
y_* \sim 2 \exp\Big[-\frac{1}{2 {\epsilon}^2} - \frac{l}{ \epsilon} \Big].
\end{equation}
Figure 7 depicts the branch iii connnector for $\epsilon = 0.3$.  The branch iii connector has a singular 
structure as $\epsilon \rightarrow 0$.  In particular, its arclength diverges like
\begin{equation}
L \sim \frac{1}{\epsilon} + l.  
\end{equation}
Its geometric action is asymptotic to
\begin{equation}
 S \sim \epsilon.
\end{equation}

In the limit $\epsilon \rightarrow 0$ with $l$ fixed, we see that the branch iii connector is the most 
probable path, beating out connector i, for $0 < l < l_c$.  The ``strategy" of of connector iii is clear:  
Descend from $y = 1$ to the $x$-axis, stay close to the $x$-axis for a long time, and finally ascend back 
up to $y = 1$.   In this way, the resistance against the deterministic flow is minimized.  The descent and 
ascent branches have little cost, because they are almost parallel to the deterministic flow.  Most of the 
leftwards motion happens
along the segment close to the $x$-axis, where the headwind is small.  

We expect that as we close the gap between $\mathbf{a}$ and $\mathbf{b}$ by decreasing $l$, we'll 
eventually find $l = l_*$ so that the branch i connector becomes the most probable path for $0 < l < 
l_*$.  By an elementary calculation, we find $l_* \sim \epsilon$ as $\epsilon \rightarrow 0$.


\section{Conclusions and Future Directions}
\label{sec:summary}

The breaking of detailed balance is ``made visible" by the splitting of forward and backward most 
probable paths between two fixed endpoints.  This is a first hint that
detailed balance and its breaking can be inferred directly from recorded histories of a stochastic 
dynamical system, in a suitable space of observables.  A first ``obvious" proposal, while achievable in 
principle, may often be difficult to implement in practice:  this involves collecting records of trajectories 
that connect two small regions around given endpoints, and observing directly the aforementioned 
splitting between forward and backward paths.
At small noise levels, there are long waits to collect these trajectories, and then you might want to have 
many of them for averaging.  It would be much better if detailed balance or its breaking can be detected 
by some simple processing of data from a few, or even one trajectory. Such a proceedure in fact exists.  
It has been fully developed by the authors for linear stochastic dynamical systems, and numerically 
tested for a simple circuit example \cite{Ghanta_PRE_2017}.  More recently, we've found that its 
nonlinear generalization is straightforward and that story will get its own paper 
\cite{Neu_AreaTensorNonlinear_2017}.


\section{acknowledgement}
We gratefully acknowledge the assistance of Varun Gudapati with the preparation of Figures 1, 2, and 3.


\end{document}